\documentclass[aps,prl,twocolumn,superscriptaddress,floatfix,nobibnotes]{revtex4-2}

\usepackage[utf8]{inputenc}
\usepackage[T1]{fontenc}
\usepackage{amsmath,amssymb,bm}
\usepackage{graphicx}
\usepackage{xcolor}
\usepackage{hyperref}
\usepackage{siunitx}
\usepackage{braket}

\hypersetup{
  colorlinks=true,
  linkcolor=blue,
  citecolor=blue,
  urlcolor=blue
}

\begin{document}

\title{Full Schmidt characterization of spatiotemporally entangled states produced from spontaneous parametric down-conversion}
 
\author{Rakesh Pradhan}
\author{Girish Kulkarni}
\email{girishk@iitrpr.ac.in}
\affiliation{Department of Physics, Indian Institute of Technology Ropar, Rupnagar, Punjab 140001, India}


\date{\today}

\begin{abstract}
The full Schmidt decomposition of spatiotemporally entangled states generated from spontaneous parametric down-conversion (SPDC) has not been carried out until now due to the immense computational complexity arising from the large dimensionalities of the states. In this Letter, we utilize the rotational symmetry of the states to reduce the complexity by at least four orders of magnitude and carry out the decomposition to reveal the precise forms of the spatiotemporal Schmidt modes and the Schmidt spectrum spanning over $10^4$ modes. We show that the Schmidt modes have a phase profile with a transverse spatial vortex structure that endows them with orbital angular momentum at all frequencies. In the high-gain regime, these Schmidt modes broaden and the Schmidt spectrum narrows with increasing pump strength. Our work can spur novel applications at the intersection of quantum imaging and spectroscopy that utilize entangled states produced from SPDC.

\end{abstract}

\maketitle


It is known that the photons generated from spontaneous parametric down-conversion (SPDC) are spatiotemporally entangled, i.e, entangled across the transverse spatial and temporal degrees of freedom (DoFs) simultaneously \cite{barreiro2005prl}. These DoFs separately already serve as excellent platforms for encoding high-dimensional quantum states \cite{erhard2020natrevphys}, which are known to have important advantages such as improved security and noise resilience in communication protocols \cite{karimipour2002pra,cerf2002prl,fujiwara2003prl,ecker2019prx}, efficiency in implementations of computing \cite{ralph2007pra,lanyon2009natphys,reimer2019natphys}, enhanced sensitivity in metrology schemes \cite{jha2011pra2,dambrosio2013natcomm}, and improved noise robustness in experimental tests on quantum foundations \cite{kaszli2000prl,collins2002prl}. However, the entanglement between the two DoFs can provide a crucial pathway to multiplicatively increase the dimensionality of the encoding through a tensor product of their Hilbert spaces \cite{xie2015natphot,graham2015natcomm,zhao2019prl,liu2022prl,zeitler2022prapp,levy2024opticaquantum}. Consequently, researchers are now turning their attention towards characterizing spatiotemporally entangled states produced from SPDC with the aim of harnessing them for quantum applications \cite{gatti2009prl,jedrkiewicz2012prl,jedrkiewicz2012prl2,osorio2008njp,caspani2010pra, gatti2012pra, baghdasaryan2022pra, gutierrez2024pra}.

A powerful mathematical tool for characterizing bipartite pure states is the Schmidt decomposition \cite{nielsen&chuang}. In particular, the spatiotemporally entangled bipartite pure state $\Psi({\bm q}_{s},\omega_{s};{\bm q}_{i},\omega_{i})$ produced from SPDC of a spatiotemporally perfectly coherent pump field admits a singular value decomposition (SVD) of the form
\begin{align}\label{schmidt}
 \Psi({\bm q}_{s},\omega_{s};{\bm q}_{i},\omega_{i})=\sum_{k=0}^{\infty} \sqrt{\lambda_{k}}\,u_{k}(\bm{q}_{s},\omega_{s})\,v_{k}(\bm{q}_{i},\omega_{i}),
\end{align}
where $\bm{q}_{j}\equiv(q_{jx},q_{jy})$ and $\omega_{j}$ represent the transverse wavevector and angular frequency for $j=s,i$ corresponding to the signal and idler fields, respectively. The non-negative real weights $\left\{\lambda_{k}\right\}$ that satisfy $\sum_{k}\lambda_{k}=1$ constitute the Schmidt spectrum, and the functions $\left\{u_{k}(\bm{q}_{s},\omega_{s}),v_{k}(\bm{q}_{i},\omega_{i})\right\}$ represent the Schmidt modes. In the low-gain regime, these modes are populated by one photon each, whereas in the high-gain regime, these modes can be populated by higher but equal numbers of photons \cite{kulkarni2022prr}. The effective width of the Schmidt spectrum $\{\lambda_{k}\}$ given by the Schmidt number $K\equiv1/\sum_{k}\lambda^2_{k}$ quantifies the degree of entanglement in the state. Thus, the Schmidt decomposition of the entangled state reveals critical information in terms of its Schmidt modes and its Schmidt spectrum that can play an important role in its optimal deployment in quantum protocols.

Until now, the Schmidt decomposition has been limitedly carried out for the postselected temporal wavefunction $\psi(\omega_{s},\omega_{i})=\Psi({\bm q}_{s}=0,\omega_{s};{\bm q}_{i}=0,\omega_{i})$ corresponding to collinear SPDC \cite{law2000prl,law2004prl,wasilewski2006pra,christ2011njp, cui2020pra,huo2020prl,chen2021prr}, and the postselected transverse spatial wavefunction $\psi({\bm q}_{s},{\bm q}_{i})=\Psi({\bm q}_{s},\omega_{s}=\omega_{p0}/2;{\bm q}_{i},\omega_{i}=\omega_{p0}/2)$ corresponding to degenerate SPDC, where $\omega_{p0}$ is the central frequency of the pump   \cite{miatto2012epjd2,miatto2012epjd1,exter2006pra,straupe2011pra,kulkarni2017natcomm,kulkarni2018pra,averchenko2020pra,amooei2025pra,fan2025acsphotonics}. For the special case in which the pump amplitude function and the phase-matching function can both be approximated to have a Gaussian form, a closed-form analytic solution in terms of the Hermite polynomials or Laguerre polynomials is possible \cite{miatto2012epjd2}. However, in the general case, the decomposition has to be carried out numerically \cite{miatto2012epjd1}. For instance, the Schmidt decomposition of $\psi(\omega_{s},\omega_{i})$ can be numerically computed by performing the SVD of the corresponding $(N\times N)$ matrix, where $N$ is the number of sample points used to discretize each frequency variable \cite{law2000prl,law2004prl,wasilewski2006pra,christ2011njp, cui2020pra,huo2020prl,chen2021prr}. Similarly, the Schmidt decomposition of $\psi({\bm q}_{s},{\bm q}_{i})$ can be computed by performing the SVD of the corresponding $(N\times N\times N\times N)$ 4-dimensional (4D) tensor, which is quite cumbersome but still tractable for $N\sim 100$  \cite{exter2006pra,straupe2011pra,kulkarni2017natcomm,kulkarni2018pra,averchenko2020pra}. Recently, a study has exploited the quasi-homogeneity and isotropy of the generated field in order to efficiently obtain the spatial Schmidt modes in high-gain SPDC \cite{amooei2025pra}. Thus, there has been substantial work on the Schmidt characterization of the postselected temporal and spatial parts of the entangled state separately.

In contrast, the Schmidt mode characterization of the complete spatiotemporally entangled state produced from SPDC has proved to be extremely challenging and remains unachieved till date. This is because the SVD of the $(N\times N\times N\times N\times N\times N)$ 6D tensor corresponding to $\Psi({\bm q}_{s},\omega_{s};{\bm q}_{i},\omega_{i})$ is computationally cumbersome even for $N\sim100$. As a result, researchers have only been able to investigate the inseparability of the spatial and temporal DoFs through limited means such as the nonfactorable nature of the `X'-shaped spatiotemporal intensity profile \cite{gatti2009prl,jedrkiewicz2012prl,jedrkiewicz2012prl2}. Other studies have quantified the spatiotemporal Schmidt number and shown that it cannot be computed as a simple product of the spatial and temporal Schmidt numbers \cite{osorio2008njp,caspani2010pra, gatti2012pra, baghdasaryan2022pra}. While such studies do shed some light on how the correlations in one DoF affect the purity of the bipartite state in another DoF, they do not completely and uniquely characterize the entangled state and its correlations across the two DoFs. As a result, it is important to perform the Schmidt decomposition of the complete entangled state and obtain the exact form of the Schmidt modes and the Schmidt spectrum in order to optimally harness such states for quantum applications.

In this Letter, we show that in common experimental scenarios involving a rotationally symmetric pump field, the computational complexity of the Schmidt decomposition can be reduced by a factor of $\mathcal{O}(N^{2}/\log N)$, which for $N=300$ yields a speedup factor greater than $10^4$. Consequently, we are able to carry out the Schmidt decomposition for the first time and compute the precise form of the spatiotemporal Schmidt modes alongwith the Schmidt spectrum. We show that the Schmidt modes possess a phase profile that has a transverse spatial vortex structure that imparts them with orbital angular momentum (OAM) at all frequencies. The corresponding Schmidt spectrum has a large dimensionality with Schmidt number $K\sim 10^4$. Our efficient method also enables us to characterize the variation of $K$ with respect to experimental parameters such as the pump beam-waist and crystal length. We then consider the high-gain regime of SPDC and show that the Schmidt modes broaden while the Schmidt spectrum narrows with increasing pump amplitude. Our work can have important implications for quantum protocols based on entangled states produced from SPDC.

We consider type-I SPDC from a thin crystal pumped by a spatiotemporally perfectly coherent pump field. In the low-gain limit, the output two-photon wavefunction $\Psi({\bm q}_{s},\omega_s; {\bm q}_{i},\omega_i)$ can be written as
\begin{align}\notag
\Psi({\bm q}_{s},\omega_s; {\bm q}_{i},\omega_i)
&= C \int\!\!\!\int\!\!\!\int\!\!\!\int \mathrm{d}\omega_s\, \mathrm{d}\omega_i\, \mathrm{d}{\bm q}_{s}\,\mathrm{d}{\bm q}_{i}\\\label{wavefunction}&\hspace{-5mm}\times A_{p}({\bm q}_{p},\omega_p)
\,\text{sinc}\left(\frac{\Delta k_z L}{2}\right) e^{i \Delta k_z L / 2},
\end{align}
where $C$ is a constant, $\mathrm{sinc}(x)\equiv (\sin x)/x$, $\Delta k_{z}\equiv k_{pz}-k_{sz}-k_{iz}$ is the longitudinal wavevector mismatch, and $L$ is the crystal length. Here, $A_{p}({\bm q}_{p},\omega_p)$ denotes the pump Fourier amplitude in terms of the pump transverse wavevector ${\bm q}_{p}$ and pump frequency $\omega_{p}$, which satisfy ${\bm q}_{p}={\bm q}_{s}+{\bm q}_{i}$ and $\omega_p=\omega_{s}+\omega_{i}$ owing to transverse momentum and energy conservation, respectively.

If the pump field is rotationally symmetric, then $A_{p}({\bm q}_{p},\omega_p)=A_{p}(|{\bm q}_{s}+{\bm q}_{i}|,\omega_s+\omega_i)$. In addition, if transverse walk-off and astigmatism effects are ignored \cite{walborn2010pr,karan2020jo}, then $\Delta k_{z}=\Delta k_{z}(|{\bm q}_{s}+{\bm q}_{i}|,|{\bm q}_{s}|,|{\bm q}_{i}|,\omega_{s},\omega_{i})$. Under these conditions, Eq.~(\ref{wavefunction}) implies that $\Psi({\bm q}_{s},\omega_s; {\bm q}_{i},\omega_i)$ is a function of $|{\bm q}_{s}|,|{\bm q}_{i}|,\omega_{s},\omega_{i}$, and $|{\bm q}_{s}+{\bm q}_{i}|$. In other words, if we denote ${\bm q}_{j}\equiv(q_{j},\phi_{j})$ in polar coordinates for $j=s,i$ and $\Delta\phi_{si}\equiv(\phi_{s}-\phi_{i})$, then because $|{\bm q}_{s}+{\bm q}_{i}|=\sqrt{q^2_{s}+q^2_{i}+2q_{s}q_{i}\cos\Delta\phi_{si}}$, it follows that
\begin{align}\label{rotsym}
 \Psi({\bm q}_{s},\omega_s; {\bm q}_{i},\omega_i)\equiv\Psi(q_{s},\omega_{s},q_{i},\omega_{i},\Delta\phi_{si}).
\end{align}
In other words, the output state is rotationally symmetric because it depends only on the angular separation $\Delta\phi_{si}\equiv(\phi_{s}-\phi_{i})$ instead of both angles $\phi_{s}$ and $\phi_{i}$ individually. Thus, $\Psi(q_{s},\omega_{s},q_{i},\omega_{i},\Delta\phi_{si})$ is a function of only five independent variables instead of six. Using Fourier's theorem, one can then define
\begin{align}\notag
 \alpha_{l}(q_{s},\omega_{s},q_{i},\omega_{i})&=\frac{1}{2\pi}\int_{0}^{2\pi}\mathrm{d}\Delta\phi_{si}\,\Psi(q_{s},\omega_{s},q_{i},\omega_{i},\Delta\phi_{si})\\\label{fourier}&\hspace{3cm}\times e^{-il\Delta\phi_{si}},
\end{align}
where $\alpha_{l}(q_{s},\omega_{s},q_{i},\omega_{i})$ for integer $l$ denote Hilbert-Schmidt kernels with decompositions of the form
\begin{align}\notag
 \alpha_{l}(q_{s},\omega_{s},q_{i},\omega_{i})&=\sum_{m=0}^{\infty} \sqrt{\lambda_{lm}}\,u_{lm}(q_{s},\omega_{s})\\\label{schmidt2}&\hspace{3cm}\times v_{lm}(q_{i},\omega_{i}).
\end{align}
Upon combining equations (1),(3),(4) and (5), one finds
\begin{align}\notag
 \Psi(q_{s},\omega_{s},q_{i},\omega_{i},\Delta\phi_{si})&=\sum_{l=-\infty}^{+\infty}\sum_{m=0}^{\infty} \sqrt{\lambda_{lm}}\\\label{schmideqn}&\hspace{-2cm}\times\,u_{lm}(q_{s},\omega_{s})\,e^{il\phi_{s}}\,\,v_{lm}(q_{i},\omega_{i})\,e^{-il\phi_{i}}.
\end{align}
Comparing the above equation with Eq.~(\ref{schmidt}), it is evident that $\{\lambda_{lm}\}$ constitute the Schmidt spectrum, whereas the functions $\{u_{lm}(q_{s},\omega_{s})\,e^{il\phi_{s}}\}$ and $\{v_{lm}(q_{i},\omega_{i})\,e^{-il\phi_{i}}\}$ represent the Schmidt modes. Thus, the Schmidt modes have a vortex phase structure in the transverse spatial plane for all frequencies. The integer index $l$ labeling each Schmidt mode denotes the vortex winding number and quantifies the OAM per photon in units of $\hbar$. The modes have pair-wise opposite OAM values $l$ and $-l$ for the signal and idler fields as expected from the conservation of OAM. The index $m$ takes only non-negative integer values, but we are not aware of its physical interpretation. The exchange symmetry $\Psi(q_{s},\omega_{s},q_{i},\omega_{i},\phi_{s}-\phi_{i})=\Psi(q_{i},\omega_{i},q_{s},\omega_{s},\phi_{i}-\phi_{s})$ leads to $\lambda_{lm}=\lambda_{(-l)m}$ and $|u_{lm}(q_{j},\omega_{j})|=|u_{(-l)m}(q_{j},\omega_{j})|=|v_{lm}(q_{j},\omega_{j})|=|v_{(-l)m}(q_{j},\omega_{j})|$ for $j=s,i$. In fact, $v_{lm}(q_{j},\omega_{j})=u_{lm}(q_{j},\omega_{j})\,e^{i\beta_{lm}}$  for $j=s,i$, where $\beta_{lm}$ are constant phases (see Supplemental Material). Thus, owing to rotational symmetry, it suffices to only numerically carry out the SVD Eq.~(\ref{schmideqn}) of the 4D tensor $\alpha_{l}(q_{s},\omega_{s},q_{i},\omega_{i})$ for non-negative $l$ instead of carrying out the SVD of the full 6D tensor $\Psi({\bm q}_{s},\omega_s; {\bm q}_{i},\omega_i)$.
\begin{figure}[t]
	\includegraphics[scale=0.41]{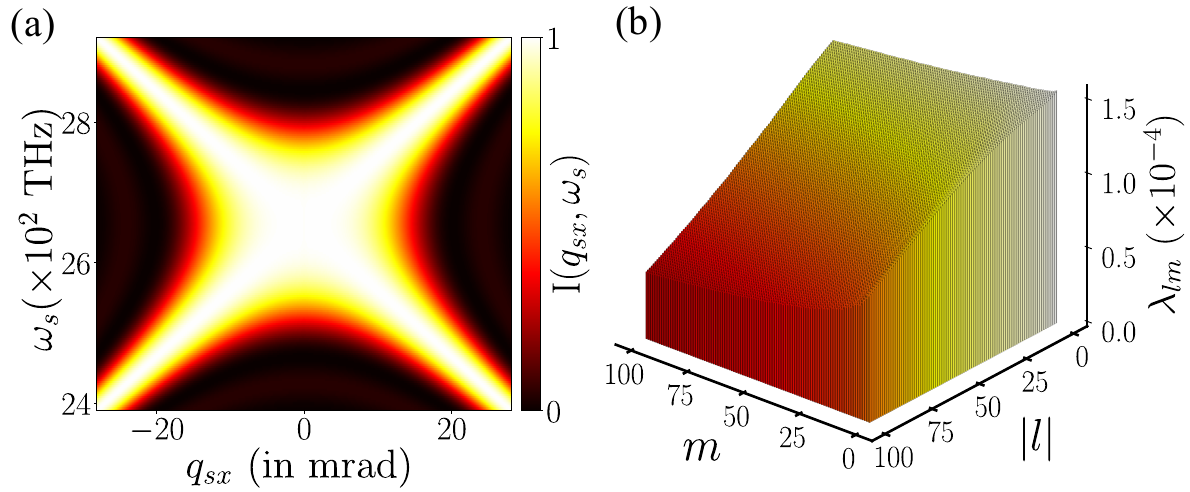}\vspace{-3mm}
	\caption{Low-gain SPDC: (a) Spatiotemporal intensity profile $I(q_{sx},\omega_{s})=G^{(1)}(q_{sx},\omega_{s};q_{sx},\omega_{s})=\sum_{l,m}\lambda_{lm}|u_{lm}(q_{sx},\omega_{s})|^2$, and (b) Spatiotemporal Schmidt spectrum $\left\{\lambda_{lm}\right\}$.}\vspace{-3mm}
    \label{fig1}
\end{figure}

Our procedure for numerically computing the decomposition can then be described as follows: (i) We first compute $\alpha_{l}(q_{s},\omega_{s},q_{i},\omega_{i})$ from $\Psi(q_{s},\omega_{s},q_{i},\omega_{i},\phi_{s}-\phi_{i})$ for each non-negative $l$ through Eq.~(\ref{fourier}) using the fast Fourier transform (FFT) algorithm, which has a complexity of $\mathcal{O}(N\log N)$ \cite{cooleytukey1965moc}. (ii) We then compute the SVD Eq.~(\ref{schmidt2}) of the sparse 4D tensor corresponding to each $\alpha_{l}(q_{s},\omega_{s},q_{i},\omega_{i})$ to get $\lambda_{lm}$ and $u_{lm}$ for all $m$, which has a complexity of $\mathcal{O}(N^6)$ (see Supplemental Material for details). This implies that the cumulative complexity of the procedure is $\mathcal{O}(N^7\log N)$. In contrast, the direct SVD of the 6D tensor corresponding to $\Psi({\bm q}_{s},\omega_{s};{\bm q}_{i},\omega_{i})$ has a complexity of $\mathcal{O}(N^9)$. In other words, our procedure achieves a speedup of at least $\mathcal{O}(N^2/\log N)$, which in our case of $N=300$ already implies a speedup factor greater than $10^4$. As a result, our highly-efficient method enables us for the first time to characterize the spatiotemporal Schmidt modes and the Schmidt spectrum of the full  spatiotemporally entangled states produced from SPDC.

We note that the Schmidt number $K=1/\sum_{l,m}\lambda^2_{lm}$ can be computed even without explicitly computing the Schmidt decomposition. To this end, one can first compute the reduced density matrix or first-order correlation function $G^{(1)}({\bm q}_{s},\omega_{s};{\bm q}'_{s},\omega'_{s})$ of the signal field as (see Supplemental Material for details)
\begin{align}\notag
 G^{(1)}({\bm q}_{s},\omega_{s};{\bm q}'_{s},\omega'_{s})&=\iint \mathrm{d}{\bm q}_{i}\,\mathrm{d}\omega_{i}\,\Psi({\bm q}_{s},\omega_{s};{\bm q}_{i},\omega_{i})\\\label{reducedstate}&\hspace{1cm}\times\Psi^*({\bm q}'_{s},\omega'_{s};{\bm q}_{i},\omega_{i}),
\end{align}
which owing to rotational symmmetry satisfies $G^{(1)}({\bm q}_{s},\omega_{s};{\bm q}'_{s},\omega'_{s})\equiv G^{(1)}(q_{s},\omega_{s},q'_{s},\omega'_{s},\Delta\phi_{ss'}=\phi_{s}-\phi'_{s})$, and then evaluate the Schmidt number $K$ as (see Supplemental Material for details)
\begin{align}\notag
 K&=\Bigg[2\pi
    \iint \mathrm{d}q_{s}\,\mathrm{d}\omega_{s}
    \iint \mathrm{d}q'_{s}\,\mathrm{d}\omega'_{s}
    \int \mathrm{d}\Delta\phi_{ss'}\Bigg.\\&\Bigg.\hspace{1cm}\times\,\left|G^{(1)}(q_{s},\omega_{s},q'_{s},\omega'_{s},\Delta\phi_{ss'})\right|^{2}\Bigg]^{-1}.
\end{align}
\begin{figure*}[t]
    \centering
    \includegraphics[width=175mm, height=76mm]{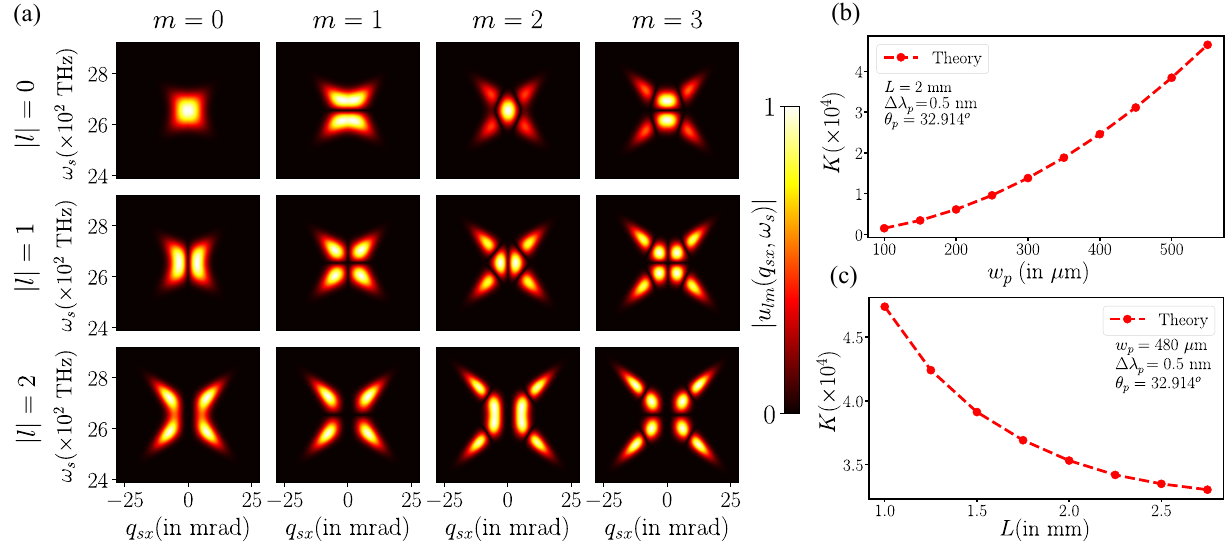}\vspace{-2mm}
    \caption{Low-gain SPDC: (a) Magnitude profiles $|u_{lm}(q_{sx},\omega_{s})|$ of the Schmidt modes for $|l|=0,1,2$ and $m=0,1,2,3$. (b) and (c) depict the variation of the Schmidt number $K$ with pump beam-waist $w_{p}$ and crystal length $L$, respectively.}
    \label{fig2}
\end{figure*}

We now present our calculations for low-gain type-I SPDC in a $\beta$-barium borate (BBO) crystal from a perfectly coherent pump field with a Gaussian spatiotemporal profile. In this case, $A_{p}({\bm q}_{p},\omega_p)$ and $\Delta k_{z}$ can be written as
\begin{subequations}\label{pumpdeltakz}
\begin{align}
A_{p}(q_p,\omega_p)&=\exp\left[ -\frac{q_p^2w_p^2}{4}\right] \exp\left[-\frac{(\omega_{p}-\omega_{p0})^2}{(\Delta \omega_{p})^2}\right],\\
 \Delta k_{z}&=\sqrt{k_{p}^2-q^2_{p}}-\sqrt{k_{s}^2-q^2_{s}}-\sqrt{k_{i}^2-q^2_{i}},
\end{align}
\end{subequations}
where $w_{p}$ is the pump beam-waist, $\Delta \omega_{p}$ is the pump spectral bandwidth, $k_{p}=\eta_{p}(\theta_{p})\omega_{p}/c$, $\eta_{p}(\theta_{p})=n_{pe}n_{po}/\sqrt{n_{po}^2\sin^2\theta_{p}+n_{pe}^2\cos^2\theta_{p}}$ is the effective refractive index experienced by the pump inside the crystal, $\theta_{p}$ is the angle between the extraordinarily polarized pump's propagation direction inside the crystal and its optic axis, $n_{p(e)o}$ is the (extra)ordinary index at the pump central wavelength $\lambda_{p0}=2\pi c/\omega_{p0}$, $k_{j}=n_{jo}(\lambda_{j})\omega_{j}/c$, where $n_{jo}(\lambda_{j})$ is the ordinary refractive index at wavelength $\lambda_{j}=2\pi c/\omega_{j}$ for $j=s,i$ corresponding to the signal and idler fields, respectively, and $c$ is the speed of light in vacuum. These quantities can be obtained using the Sellmeier relations for BBO \cite{eimerl1987jap}.
We then choose $N=300$, $\lambda_{p0}=355$ nm, $w_{p}=480$ $\mu$m, $L=2$ mm, $\theta_{p}=32.914^{\circ}$, $\Delta\lambda_{p}=\Delta\omega_{p}\lambda_{p0}/\omega_{p0}=0.5$ nm, plug the relations (\ref{pumpdeltakz}) into Eq.~(\ref{wavefunction}), and evaluate the Schmidt decomposition following our outlined procedure.

In Figs.~\ref{fig1}(a) and \ref{fig1}(b), we depict the spatiotemporal intensity $I(q_{sx}, \omega_{s})=G^{(1)}(q_{sx},\omega_{s};q_{sx},\omega_{s})=\sum_{l,m} \lambda_{lm}|u_{lm}(q_{sx},\omega_{s})|^2$ with the characteristic X-profile and the Schmidt spectrum $\left\{\lambda_{lm}\right\}$ for $|l|=[0,100]$ and $m=[0,100]$, respectively. In Fig.~\ref{fig2}(a), we represent the spatiotemporal magnitude profiles $|u_{lm}(q_{sx},\omega_{s})|$ of the first few Schmidt modes corresponding to $|l|=0,1,2$ and $m=0,1,2$ and $3$. Because of the vortex $e^{il\phi_{s}}$ in their transverse spatial phase profile, the modes for non-zero $l$ have an intensity node at $q_{sx}=0$ for all $\omega_{s}$. Also, due to the rotational symmetry of $|u_{lm}|$, it suffices to depict our plots with respect to $q_{sx}$. We show that in general, these profiles cannot be factorized into separate spatial and temporal parts (see Supplemental Material for details). In Figs.~\ref{fig2}(b) and \ref{fig2}(c), we depict the Schmidt number $K$ increasing with increasing pump beam-waist $w_{p}$, and $K$ decreasing with increasing crystal length $L$, respectively. The variation of $K$ with pump spectral bandwidth $\Delta\lambda$ and crystal angle $\theta_{p}$ are presented in the Supplemental Material. Our results constitute the first complete characterization of spatiotemporally entangled states produced from low-gain SPDC with dimensionalities of the order of $10^4$.

Next, we consider the problem of evaluating the Schmidt decomposition in the high-gain regime. In this regime, there is no known closed-form analytic expression for the wavefunction $\Psi({\bm q}_{s},\omega_s; {\bm q}_{i},\omega_i)$, but there is one for the first-order correlation function $G^{(1)}({\bm q}_{s},\omega_{s};{\bm q}'_{s},\omega'_{s})$ of the signal field. It is still possible to find the decomposition because equations (\ref{schmideqn}) and (\ref{reducedstate}) imply that
\begin{align}\notag
f_{l}(q_{s},\omega_{s},q'_{s},\omega'_{s})&=\frac{1}{2\pi}\int_{0}^{2\pi}\mathrm{d}\Delta\phi_{ss''}\\\label{highgaing1}&\hspace{-1cm}\times\,G^{(1)}(q_{s},\omega_{s},q'_{s},\omega'_{s},\Delta\phi_{ss'})\,e^{-il\Delta\phi_{ss'}},
\end{align}
where $f_{l}(q_{s},\omega_{s},q'_{s},\omega'_{s})$ for integer $l$ has the coherent-mode decomposition
\begin{align}
 f_{l}(q_{s},\omega_{s},q'_{s},\omega'_{s})=\sum_{m=0}^{\infty} \lambda_{lm}\,u_{lm}(q_{s},\omega_{s})\,u^*_{lm}(q'_{s},\omega'_{s}).
\end{align}
Thus, the decomposition procedure in the high-gain regime is as follows: (i) We first use the Fourier relation (\ref{highgaing1}) to obtain $f_{l}(q_{s},\omega_{s},q'_{s},\omega'_{s})$ for each non-negative $l$, and (ii) Diagonalize the 4D tensor corresponding to $f_{l}(q_{s},\omega_{s},q'_{s},\omega'_{s})$ to obtain $\lambda_{lm}$ and $u_{lm}(q_{s},\omega_{s})$ for all $m$. Again, this procedure is faster by a factor of $\mathcal{O}(N^2/\log N)$ compared to diagonalizing the 6D tensor associated to $G^{(1)}({\bm q}_{s},\omega_{s};{\bm q}'_{s},\omega'_{s})$ directly.

\begin{figure}[t]
	\includegraphics[scale=0.47]{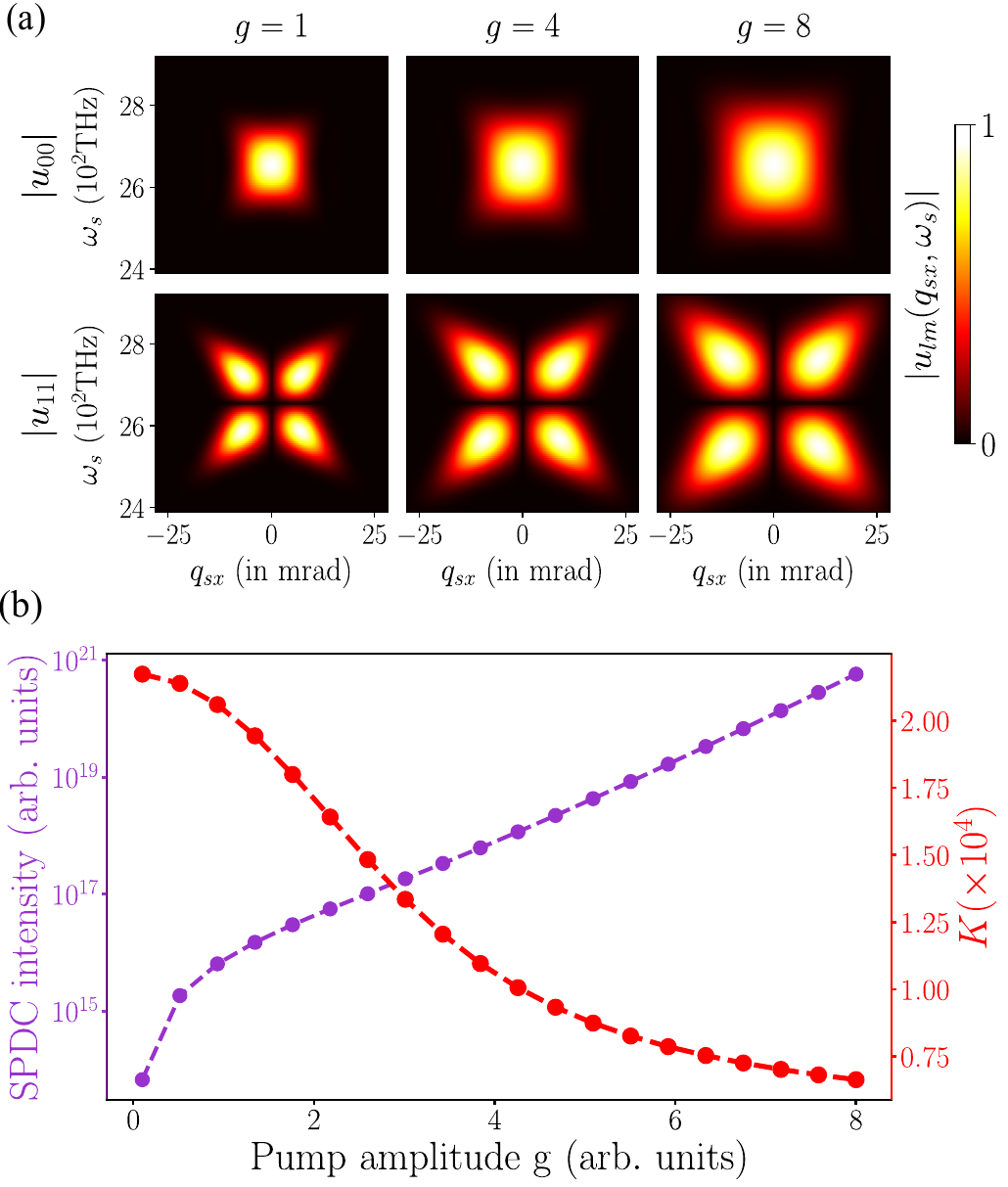}\vspace{-3mm}
	\caption{High-gain SPDC: (a) depicts the broadening of the Schmidt modes $|u_{00}(q_{sx},\omega_{s})|$ and $|u_{11}(q_{sx},\omega_{s})|$ with increasing gain for $g=1,4,$ and $8$ . (b) depicts the total integrated intensity of the signal field (left y-axis in violet) and Schmidt number $K$ (right y-axis in red) with increasing $g$.}
    \label{fig3}
\end{figure}

We now present our results for the high-gain regime wherein $G^{(1)}(\bm{q}_{s},\omega_{s};\bm{q'}_{s},\omega'_{s})$ takes the form \cite{kulkarni2022prr}
\begin{align}\notag
 G^{(1)}(\bm{q}_{s},\omega_{s};\bm{q'}_{s},\omega'_{s})&=\frac{C_{1}}{k_{sz}k'_{sz}}\iint\mathrm{d}{\bm \rho}\,\mathrm{d}t\,\langle|V_{p}({\bm \rho},t)|^2\rangle\\\notag&\hspace{-3cm}\times e^{-i[({\bm q_{s}}-{\bm q'_{s}})\cdot{\bm \rho}-(\omega_{s}-\omega'_{s})t]}\left[\frac{\mathrm{sinh}\,\Gamma(\Delta \bar{k}_{z},{\bm \rho},t)L}{\Gamma(\Delta \bar{k}_{z},{\bm \rho},t)}\right]\\\label{hgpdccorrfunc}&\hspace{-2cm}\times\left[\frac{\mathrm{sinh}\,\Gamma(\Delta \bar{k'}_{z},{\bm \rho},t)L}{\Gamma(\Delta \bar{k'}_{z},{\bm \rho},t)}\right] e^{i(\Delta \bar{k}_{z}-\Delta \bar{k}'_{z})L/2},
\end{align}
where $C_{1}$ is an overall constant, $V_{p}({\bm \rho},t)$ is the pump amplitude profile as a function of transverse position vector ${\bm \rho}$ and time $t$, $\Delta \bar{k}_{z}$ is the central value of $\Delta k_{z}$ evaluated under the conditions ${\bm q_{s}}+{\bm q_{i}}=0$ and $\omega_{s}+\omega_{i}=\omega_{p0}$, as defined in Ref.~ \cite{kulkarni2022prr}. In the rest of the paper, any quantity appearing with an overbar notation must be interpreted similarly. The quantity $\Gamma(\Delta \bar{k}_{z},{\bm \rho},t)$ takes the form \cite{kulkarni2022prr}
\begin{align}\label{gammaexp}
\Gamma(\Delta \bar{k}_{z},{\bm \rho},t)\equiv\left[\frac{C_{2}}{k_{sz}\bar{k}_{iz}}|V_{p}({\bm \rho},t)|^2-\left(\frac{\Delta \bar{k}_{z}}{2}\right)^2\right]^{1/2},
\end{align}
where $C_{2}$ is a scaling factor. We substitute
\begin{subequations}
\begin{align}
 V_{p}({\bm \rho},t)&=g\exp\left\{-\frac{\rho^2}{w^2_{p}}\right\}\exp\left\{-\frac{t^2}{2\Delta t^2}\right\},\\
\Delta \bar{k}_{z}&=k_{p}-\sqrt{k_{s}^2-q_{s}^2}-\sqrt{\bar{k}_{i}^2-q_{s}^2},
\end{align}
\end{subequations}
where $g$ is the pump amplitude and $\Delta t$ is the temporal width of the pump pulse. The values of $C_{1}$ and $C_{2}$ are appropriately chosen such that $g<<1$ corresponds to the low-gain regime and $g>1$ corresponds to the high-gain regime \cite{kulkarni2022prr}. In Fig.~\ref{fig3}(a), we depict the magnitude profiles $|u_{00}(q_{sx},\omega_{s})|$ and $|u_{11}(q_{sx},\omega_{s})|$ for $g=1,4,8$, wherein it is evident that the Schmidt modes broaden with increasing pump amplitude. In Fig.~\ref{fig3}(b), we depict the total integrated signal field intensity on the left $y$-axis in violet and the Schmidt number $K$ on the right $y$-axis in red with increasing pump amplitude. Note the onset of exponential growth of signal intensity for $g>1$ following which $K$ decreases, thereby indicating a narrowing Schmidt spectrum with increasing pump strength.

In summary, we demonstrate an efficient method to carry out the Schmidt decomposition of the spatiotemporally entangled state produced from SPDC. This method utilizes the rotational symmetry of the generated state to reduce the complexity of the decomposition by $\mathcal{O}(N^2/\log N)$, which for $N=300$ corresponds to at least four orders of magnitude. As a result, we are able to compute the precise forms of the spatiotemporal Schmidt modes and the corresponding Schmidt spectrum for the first time. We show that the Schmidt modes have a phase profile with a transverse spatial vortex structure that imparts them with OAM at all frequencies. We also quantify the variation of the Schmidt number with respect to experimental parameters such as pump beam-waist and crystal length. Finally, we show that in the high-gain regime, the Schmidt modes broaden and the Schmidt spectrum narrows with increasing pump strength.

Our work can have direct implications for experimental schemes in quantum imaging \cite{brida2010natphot, cameron2024science,defienne2024natphot} and spectroscopy \cite{matsuzaki2022natcomm, dorfman2016rmp}, wherein the precise knowledge about the spatiotemporal Schmidt modes and the Schmidt spectrum can aid in optimizing the resolution and sensitivity of those schemes through the judicious choice of detection modes. In addition, our work can pave the way towards the use of OAM-carrying spatiotemporal modes for realizing communication protocols with higher channel capacity \cite{willner2015aop}. Moreover, our technique can also be adapted to efficiently compute the spatiotemporal coherent eigenmodes of rotationally symmetric spatiotemporally partially coherent fields \cite{hyde2023jo}. Consequently, our technique can be useful for investigating space-time coupling in multimode fibers and lasers \cite{wright2015natphot,xiong2016prl,chriki2018pra}. We finally hope that our work finds other applications in the wider research landscape of spatiotemporal light fields \cite{shen2023jo,liu2024natcomm,zhan2024aop}.

\begin{acknowledgments}
We acknowledge financial support through the National Quantum Mission (NQM) of the Department of Science and Technology, Government of India and the initiation research grant received
from IIT Ropar.
\end{acknowledgments}

\bibliographystyle{apsrev4-2}
\bibliography{spatiotemporalnew}

\end{document}



\makeatletter
\def\@seccntformat#1{\csname the#1\endcsname.\quad}
\makeatother

\renewcommand{\thesection}{\Roman{section}}
\renewcommand{\thesubsection}{\Roman{section}.\arabic{subsection}}

\setcounter{secnumdepth}{3}


\title{Supplemental Material: \textrm{Full Schmidt characterization of spatiotemporally entangled states produced from
spontaneous parametric down-conversion
}}

\author{Rakesh Pradhan}
\author{Girish Kulkarni}
\email{girishk@iitrpr.ac.in}
\affiliation{Department of Physics, Indian Institute of Technology Ropar, Rupnagar, Punjab 140001, India}

\date{\today}

\maketitle

\vspace{1em}

\section{Array Flattening Method}

In order to obtain the two-dimensional (2D) spatiotemporal modal functions $u_{lm}(q_s,\omega_{s})$, we must perform the singular value decomposition (SVD) of the full four-dimensional azimuthal kernel $\alpha_l(q_s,\omega_{s};q_i,\omega_i)$ of Eq.~(5) of the main paper. As the numerical algorithm for SVD is defined for a 2D tensor (matrix), the four-dimensional kernel is first bijectively reshaped into a matrix using an array-flattening procedure. Each pair of variables is mapped onto a single composite index,
\begin{equation}
(q_s,\omega_s) \;\rightarrow\; \mu, \qquad
(q_i,\omega_i) \;\rightarrow\; \nu ,
\end{equation}
so that the kernel is represented as a matrix $A^{(l)}_{\mu\nu}$.

\begin{wrapfigure}{r}{0.4\textwidth}
    \centering
    \vspace{-10pt} 
    \includegraphics[width=0.45\textwidth]{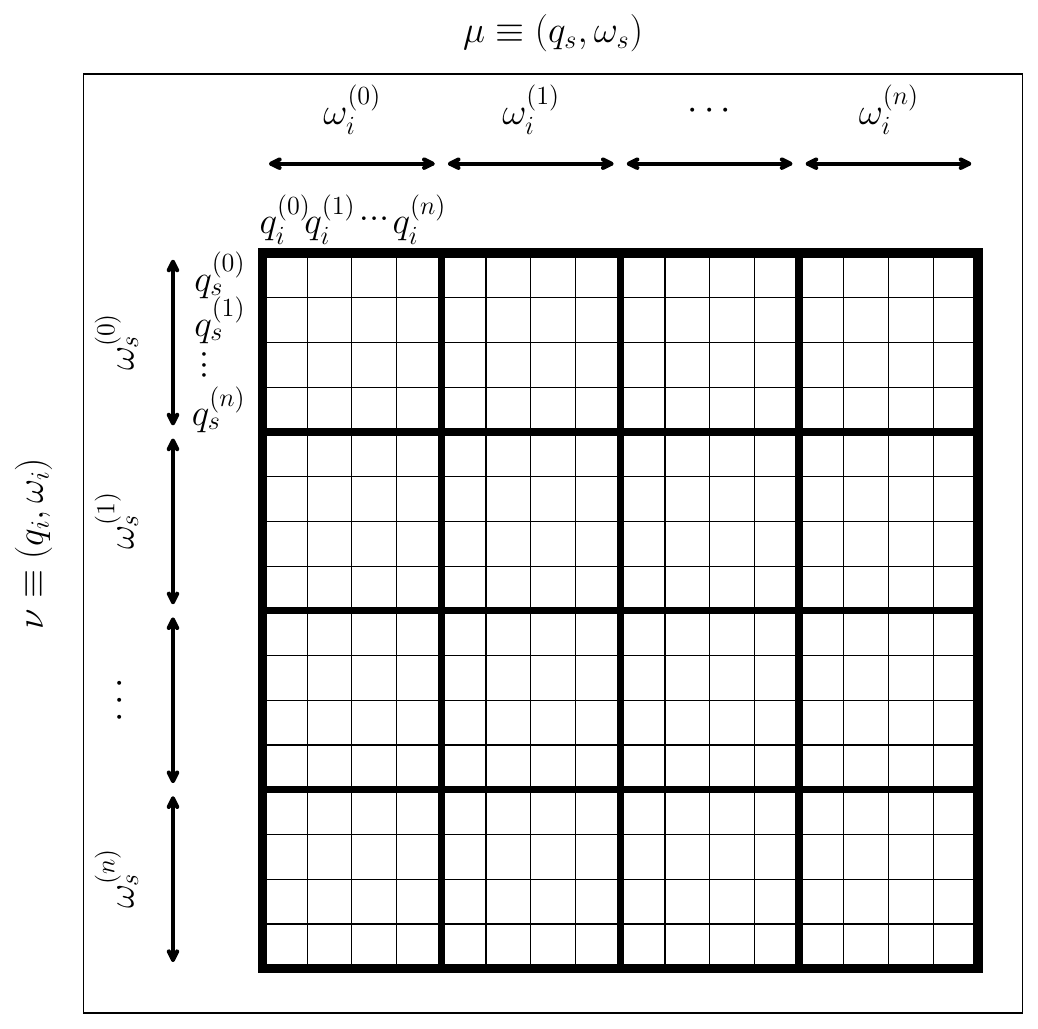}
    \caption{Schematic depiction of the flattening procedure used to reshape the 4D tensor $\alpha_l(q_s, \omega_s;  q_i, \omega_i)$ into a 2D matrix.}
    \label{smfig1}
    \vspace{3pt} 
\end{wrapfigure}

Each of the variables $(q_s,\omega_s,q_i,\omega_i)$ is discretized using $N$ grid points. After discretization, the kernel contains $N^4$ elements and is reshaped into a square matrix of dimension $N^2 \times N^2$,

\begin{equation}
A^{(l)} \in \mathbb{C}^{N^2 \times N^2}.
\end{equation}

This matrix exhibits a natural block structure: It can be viewed as an $N \times N$ array of blocks, where each block corresponds to a fixed pair of temporal coordinates $(\omega_s,\omega_i)$. Each block is itself an $N \times N$ matrix containing the values of $\alpha_l(q_s,\omega_{s};q_i,\omega_i)$ over all spatial coordinates $(q_s, q_i)$.

A singular-value decomposition of the flattened matrix takes the form,
\begin{equation}
A^{(l)} = U^{(l)} \Sigma^{(l)} \bigl(V^{(l)}\bigr)^\dagger ,
\end{equation}
which yields the Schmidt spectrum given by the singular values in $\Sigma^{(l)}$, as well as the corresponding flattened Schmidt modes.

Because the array-flattening procedure is bijective and preserves all correlations, the total number of modes and their singular values remain unchanged under this projection. Consequently, the Schmidt spectrum obtained from the flattened two-dimensional matrix is identical to that of the original four-dimensional spatiotemporal kernel. The flattened Schmidt modes can be reshaped back into functions of $(q_j,\omega_j)$ to recover the actual two-dimensional spatiotemporal modal functions $u_{lm}(q_j,\omega_j)$.\\

The computational complexity of performing a singular value decomposition (SVD) of a dense \(N\times N\) matrix scales as \(\mathcal{O}(N^{3})\). Consequently, a direct decomposition of a six-dimensional tensor flattened into a \((N^{3}\times N^{3})\) matrix leads to a computational cost of \(\mathcal{O}(N^{9})\). In our approach, we first extract the coefficients \(\alpha_l(q_s,\omega_s,q_i,\omega_i)\) using the fast Fourier Transform (FFT) algorithm, which has a complexity of \(\mathcal{O}(N\log N)\). The resulting \(\alpha_l\) forms a four-dimensional tensor, equivalently a flattened \((N^{2}\times N^{2})\) matrix, whose SVD scales as \(\mathcal{O}(N^{6})\). Thus, our method reduces the overall computational complexity from \(\mathcal{O}(N^{9})\) to \(\mathcal{O}(N^{7}\log N)\), yielding a theoretical speedup of order \(\mathcal{O}(N^{2}/ \log N)\). Moreover, the matrix $A^{(l)}$ is sparse because its elements are nonzero only when the phase-matching conditions of energy and momentum conservation are satisfied. This sparsity further reduces the effective complexity of the decomposition, which depends on the number of nonzero elements, leading to an additional improvement in computational efficiency beyond the dense-matrix scaling. As a result, the actual speedup achieved is usually much larger than $\mathcal{O}(N^{2}/\log N)$.

\section{Properties of Schmidt Modes}
We note that
\begin{align}
\alpha_l(q_s, \omega_s; q_i, \omega_i)
=
\frac{1}{2\pi}
\int_0^{2\pi}
d\Delta \phi_{si}\,
\Psi(q_s, \omega_s; q_i, \omega_i, \Delta \phi_{si})\,
e^{i l\Delta \phi_{si}} .
\label{azimuthal-kernel}
\end{align}

The exchange of the signal and idler variables corresponds to
\[
\Psi(q_s, \omega_s; q_i, \omega_i, \Delta \phi_{si})
\;\longrightarrow\;
\Psi(q_i, \omega_i, q_s,\omega_s, -\Delta \phi_{si}) .
\]
For type-I SPDC, the signal and idler photons have identical polarization, and the biphoton wavefunction is symmetric under their exchange. Consequently,
\begin{equation}
\Psi( q_i, \omega_i, q_s, \omega_s, -\Delta \phi_{si})
=
\Psi(q_s, \omega_s, q_i,  \omega_i,  \Delta \phi_{si})
=
\Psi(q_s, \omega_s, q_i, \omega_i,  |\Delta \phi_{si}|) ,
\label{exchange-symmetry}
\end{equation}
which implies that $\Psi$ is an even function of $\Delta \phi_{si}$, i.e,
\[
\Psi(q_s, \omega_s; q_i, \omega_i, \Delta \phi_{si}) = \Psi(q_s, \omega_s; q_i, \omega_i, -\Delta \phi_{si}) .
\]

We now evaluate the kernel with opposite azimuthal index:
\begin{align}
\alpha_{-l}(q_s, \omega_s, q_i, \omega_i )
&=
\frac{1}{2\pi}
\int_0^{2\pi}
d\Delta \phi_{si}\,
\Psi(q_s, \omega_s; q_i, \omega_i, \Delta \phi_{si})\,
e^{-i l\Delta \phi_{si}} .
\end{align}
Performing the change of variables $\Delta \phi_{si} \rightarrow -\Delta \phi_{si}$, we obtain
\begin{align}
\alpha_{-l}(q_s, \omega_s, q_i, \omega_i )
&=
\frac{1}{2\pi}
\int_{0}^{-2\pi}
(-d\Delta \phi_{si})\,
\Psi(q_s, \omega_s; q_i, \omega_i, -\Delta \phi_{si})\,
e^{i l\Delta \phi_{si}} \notag\\
&=
\frac{1}{2\pi}
\int_{-2\pi}^{0}
d\Delta \phi_{si}\,
\Psi(q_s, \omega_s; q_i, \omega_i, \Delta \phi_{si})\,
e^{i l\Delta \phi_{si}} \notag\\
&=
\frac{1}{2\pi}
\int_{0}^{2\pi}
d\Delta \phi_{si}\,
\Psi(q_s, \omega_s; q_i, \omega_i, \Delta \phi_{si})\,
e^{i l\Delta \phi_{si}} \notag\\
&=
\alpha_l(q_s, \omega_s, q_i, \omega_i) .
\end{align}

As a consequence, the Schmidt spectrum and mode functions exhibit the degeneracy
\begin{equation}
\lambda_{lm} = \lambda_{(-l)m}, \qquad
u_{lm}(q_s, \omega_s) = u_{(-l)m}(q_s, \omega_s), \qquad
v_{lm}(q_i, \omega_i) = v_{(-l)m}(q_i, \omega_i),
\end{equation}
up to an overall phase, reflecting the underlying exchange and rotational symmetry of the biphoton state in type-I SPDC.

\subsection*{Symmetry of the azimuthal kernels under transposition}

We now show that the azimuthal Fourier kernels are symmetric under transposition, i.e, $\alpha_l(q_s, \omega_s, q_i, \omega_i)=\alpha_l(q_i, \omega_i, q_s, \omega_s)$. From the definition of Eq.~(4) of the main paper,
\begin{align}
\alpha_l(q_s, \omega_s, q_i, \omega_i)
=
\frac{1}{2\pi}
\int_0^{2\pi}
d\Delta \phi_{si}\,
\Psi(q_s, \omega_s, q_i, \omega_i, \Delta \phi_{si})\,
e^{i l\Delta \phi_{si}} .
\label{azimuthal-kernel}
\end{align}

Taking the transpose corresponds to exchanging the signal and idler variables,
\begin{align}
\alpha_l(q_s, \omega_s, q_i, \omega_i)^T=\alpha_l(q_i, \omega_i, q_s, \omega_s)
&=
\frac{1}{2\pi}
\int_0^{2\pi}
d\Delta \phi_{si}\,
\Psi( q_i, \omega_i, q_s, \omega_s, \Delta \phi_{si})\,
e^{i l\Delta \phi_{si}} .
\end{align}

For type-I SPDC, the biphoton wavefunction is symmetric under exchange of the signal and idler photons due to their identical polarization,
\begin{equation}
\Psi(q_i, \omega_i,  q_s, \omega_s, \Delta \phi_{si})
=
\Psi( q_s, \omega_s, q_i, \omega_i, \Delta \phi_{si}) .
\label{exchange-symmetry}
\end{equation}

Substituting Eq.~\eqref{exchange-symmetry} into the expression above immediately yields
\begin{align}
\alpha_l(q_s, \omega_s, q_i, \omega_i)^T
&=
\frac{1}{2\pi}
\int_0^{2\pi}
d\Delta \phi_{si}\,
\Psi(q_s, \omega_s, q_i, \omega_i, \Delta \phi_{si})\,
e^{i l\Delta \phi_{si}} \notag\\
&=
\alpha_l(q_s, \omega_s, q_i, \omega_i) .
\end{align}

Therefore, the azimuthal kernels are symmetric matrices,
\begin{equation}\label{kernel-symmetric}
\alpha_l = \alpha_l^T 
\end{equation}

Now the singular value decomposition (SVD) of $a_l$ is written as
\begin{equation}
\alpha_l = U_l \,\Sigma_l\, V_l^\dagger ,
\label{svd}
\end{equation}
where $U_l$ and $V_l$ are unitary matrices whose columns are the signal and idler Schmidt modes,
\[
U_l = (u_{l1}, u_{l2}, \dots), \qquad
V_l = (v_{l1}, v_{l2}, \dots),
\]
and $\Sigma_l = \mathrm{diag}(\lambda_{l1}, \lambda_{l2}, \dots)$ contains the Schmidt coefficients.

Taking the transpose of Eq.~\eqref{svd} and using Eq.~\ref{kernel-symmetric}, we obtain
\begin{equation}
\alpha_l
=
\alpha_l^T
=
V_l^* \,\Sigma_l\, U_l^T .
\label{svd-transpose}
\end{equation}

Comparing Eqs.~\eqref{svd} and \eqref{svd-transpose} yields
\begin{equation}
U_l \,\Sigma_l\, V_l^\dagger
=
V_l^* \,\Sigma_l\, U_l^T .
\end{equation}

Multiplying from the left by $U_l^\dagger$ and from the right by $U_l^*$ gives
\begin{equation}
\Sigma_l \, W_l
=
W_l^\dagger \, \Sigma_l ,
\qquad
W_l \equiv U_l^\dagger V_l^* .
\label{commutation}
\end{equation}

Since $\Sigma_l$ is diagonal with non-negative entries, Eq.~\eqref{commutation} implies that $W_l$ must be diagonal within each non-degenerate Schmidt subspace. Unitarity of $W_l$ then gives
\begin{equation}
W_l = \mathrm{diag}(e^{-i\beta_{l1}}, e^{-i\beta_{l2}}, \dots) .
\end{equation}

Substituting back,
\begin{equation}
V_l^* = U_l W_l
\quad\Rightarrow\quad
v_{lm}(q_{j},\omega_{j}) = e^{-i\beta_{lm}} u_{lm}(q_{j},\omega_{j}) .
\end{equation}

Therefore, the signal and idler Schmidt modes are identical up to arbitrary phase factors, i.e,
\begin{equation}
v_{lm}(q_{j},\omega_{j}) = e^{i\beta_{lm}}\, u_{lm}(q_{j},\omega_{j}) .
\label{mode-phase}
\end{equation}

\section{Spatiotemporal Correlation Function}

Following the the equation (7) in the main text, the correlation function $G^{(1)}(q_s, \omega_s,q_s', \omega_s', \phi_s, \phi_s')$ is defined as

\begin{align}
G^{(1)}(q_s, \omega_s,q_s', \omega_s', \phi_s, \phi_s') &= \int_0^{2\pi} d\theta_i  \int d\omega_i \int q_i dq_i \Psi(q_s, \omega_s, q_i,\omega_i,  \phi_s - \phi_i) \Psi^*(q_s', \omega_s', q_i,\omega_i,  \phi_s' - \phi_i) \\
 &= \int_0^{2\pi} d\alpha \int d\omega_i  \int q_i dq_i \Psi(q_s, \omega_s, q_i, \omega_i,\alpha) \Psi^*(q_s', \omega_s', q_i, \omega_i, \alpha + (\phi_s - \phi_s')) \notag\\
 &= G^{(1)}(q_s, \omega_s, q_s', \omega_s', \Delta \phi_{ss'}) \notag, 
\end{align}

where $\Delta \phi_{ss'}  = \phi_s - \phi_s'$. This first-order correlation function admits the coherent mode decomposition
\begin{equation}
G^{(1)}(q_s, \omega_s, q_s', \omega_s', \phi_s, \phi_s')
=
\sum_{l=-\infty}^{\infty}
\sum_{m=0}^{\infty}
\lambda_{lm}\,
w_{lm}(q_s, \omega_s, \phi_s)\,
w_{lm}^*(q'_s, \omega_s', \phi_s'),
\label{G1_schmidt}
\end{equation}

where $\{w_{lm}(\omega_s, q_s, \phi_s)\}$ form an orthonormal set,

\begin{equation}\label{orthonormality}
\int d\omega_s \int q_s dq_s \int d\phi_s\;
w_{lm}^*(\omega_s, q_s, \phi_s)\,
w_{l'm'}(\omega_s, q_s, \phi_s)
=
\delta_{ll'}\delta_{mm'}.
\end{equation}

The squared Hilbert--Schmidt norm of $G^{(1)}$ is given by
\begin{equation}
\int d\omega_s \int q_s dq_s \int d\phi_s
\int d\omega_s' \int q_s' dq_s' \int d\phi_s'\;
\bigl|
G^{(1)}( q_s, \omega_s, q_s', \omega_s', \phi_s, \phi_s')
\bigr|^2 .
\label{HS_norm}
\end{equation}

Substituting Eq.~\ref{G1_schmidt} into Eq.~\ref{HS_norm} and using the orthonormality relation~\ref{orthonormality}, all cross terms vanish, yielding
\begin{equation}
\int d\omega_s \int q_s dq_s \int d\phi_s
\int d\omega_s' \int q_s' dq_s' \int d\phi_s'\;
\bigl|
G^{(1)}(\omega_s, \omega_s', q_s, q_s', \phi_s, \phi_s')
\bigr|^2
=
\sum_{l=-\infty}^{\infty}
\sum_{m=0}^{\infty}
\lambda_{lm}^2 .
\label{eq:lambda_squared}
\end{equation}

Since the first-order correlation function depends only on the azimuthal
difference $\Delta\phi_{ss'}=\phi_s-\phi_s'$, the integration variables and their measures can be transformed as
\begin{equation}
\int d\phi_s \int d\phi_s'
=
2\pi \int d\Delta\phi_{ss'} .
\end{equation}
As a result, the sum of the squared Schmidt coefficients can be written as
\begin{equation}
2\pi \int d\omega_s \int q_s dq_s
\int d\omega_s' \int q_s' dq_s' \int d\Delta \phi_{ss'}\;
\bigl|
G^{(1)}(\omega_s, \omega_s', q_s, q_s',\Delta \phi_{ss'})
\bigr|^2
=
\sum_{l=-\infty}^{\infty}
\sum_{m=0}^{\infty}
\lambda_{lm}^2 .
\label{eq:lambda_squared}
\end{equation}

If $G^{(1)}$ is normalized such that $\sum_{l,m}\lambda_{lm}=1$, the Schmidt
number is therefore given by
\begin{equation}
K =
\left[
2\pi \int d\omega_s \int q_s dq_s
\int d\omega_s' \int q_s' dq_s' \int d\Delta \phi_{ss'}\;
\bigl|
G^{(1)}(\omega_s, \omega_s', q_s, q_s',\Delta \phi_{ss'})
\bigr|^2
\right]^{-1}.
\label{eq:Schmidt_Number}
\end{equation}

\subsection*{Coherent mode decomposition of correlation function}

From the eq.(6) of main text we have, 

\begin{align}\label{SVD}
&\Psi( {q}_{s}, \omega_{s}, {q}_{i}, \omega_{i},\Delta\phi_{si})
= \sum_{l=-\infty}^\infty \sum_{r=0}^{\infty} \sqrt{\lambda_{lm}}\,u_{lm}(\omega_{s}, {q}_{s})  e^{i l\theta_s}\, 
 v_{lm}(\omega_{i}, {q}_{i})\, e^{-i l \theta_i}.
\end{align}

\begin{align}\notag
G^{(1)}(q_s, \omega_s, q_s', \omega_s', \Delta \phi_{ss'}) &= \int d\omega_i \int_0^{2\pi} d\alpha \int q_i dq_i \Psi(q_s, \omega_s, q_i, \omega_i, \alpha) \Psi^*(q_s', \omega_s', q_i, \omega_i,\alpha + \Delta \phi_{ss'}) \\ \notag 
 &= \int d\omega_i \int_0^{2\pi} d\alpha \int q_i dq_i  \sum_{l=-\infty}^\infty \sum_{m=0}^{\infty} \sum_{l'=-\infty}^\infty \sum_{m'=0}^{\infty} \sqrt{\lambda_{lm}}\sqrt{\lambda_{l'm'}}\,u_{lm}({q}_{s}, \omega_{s})\,  v_{lm}({q}_{i}, \omega_{i})\, \\ \notag
 & \hspace{7cm}\times u^*_{l'm'}( {q}_{s}', \omega_{s}')\, v^*_{l'm'}( {q}_{i}, \omega_{i})\, e^{-i l'\Delta \phi_{ss'} }\, e^{i \alpha (l-l')}\\  \notag
 &= \sum_{l=-\infty}^\infty \sum_{m=0}^{\infty} \sum_{l'=-\infty}^\infty \sum_{m'=0}^{\infty} \sqrt{\lambda_{lm}}\sqrt{\lambda_{l'm'}}\,u_{lm}({q}_{s}, \omega_{s})\,
 u^*_{l'm'}( {q}_{s}', \omega_{s}')\, e^{-i l'\Delta \phi_{ss'} } \delta_{mm'} \delta_{ll'}\\
&= \sum_{l=-\infty}^\infty \sum_{m=0}^{\infty} \lambda_{lm}\,u_{lm}({q}_{s}, \omega_{s})\,
 u^*_{lm}( {q}_{s}', \omega_{s}')\, e^{-i l\Delta \phi_{ss'}}
\end{align}

Now we multiply both sides of the above equation by
$e^{i l' \Delta \phi_{ss'}}$ and integrate with respect to $\Delta \phi_{ss'}$ as

\begin{align}\notag
 \int_0^{2\pi} d\Delta \phi_{ss'}  G^{(1)}(q_s, \omega_s, q_s', \omega_s', \Delta \phi_{ss'}) \, e^{il'\Delta \phi_{ss'}} &= \int_0^{2\pi} d \Delta \phi_{ss'} \sum_{l=-\infty}^\infty \sum_{m=0}^{\infty} \lambda_{lm}\,u_{lm}({q}_{s}, \omega_{s})\,
 u^*_{lm}( {q}_{s}', \omega_{s}')\, e^{i \Delta \phi_{ss'}(l'-l) }\\ \notag
 &=\sum_{l=-\infty}^\infty \sum_{m=0}^{\infty} \lambda_{lm}\,u_{lm}({q}_{s}, \omega_{s})\,
 u^*_{lm}( {q}_{s}', \omega_{s}')\,\delta_{ll'}\\ \notag
 &= \sum_{m=0}^{\infty} \lambda_{l'm}\,u_{l'm}({q}_{s}, \omega_{s})\,
 u^*_{l'm}( {q}_{s}', \omega_{s}')\,\\ 
 &=f_{l'}(q_s, \omega_s, q_s', \omega_s')
\end{align}

\section{Non-separability of Schmidt modes}

Consider a 2D mode $u(q_s, \omega_s)$ that lives in the tensor product Hilbert space
$\mathcal{H}_s \otimes \mathcal{H}_t$, where $\mathcal{H}_s$ and $\mathcal{H}_t$ are the Hilbert spaces associated with the spatial and temporal degrees of freedom, respectively.
The mode is said to be \emph{separable} if it can be written as
\begin{equation}
u(q_s,\omega_s) = s(q_s)\, t(\omega_s),
\end{equation}
where $s \in \mathcal{H}_s$ and $t \in \mathcal{H}_t$.

In general, $u(q_s,\omega_s)$ admits a singular-value decomposition
\begin{equation}
u(\omega_s,q_s) 
=
\sum_{j=1}^{\infty}
\sqrt{\sigma_j}\,
s_j(q_s)\, t_j(\omega_s),,
\end{equation}
where $\{s_j\}$ and $\{t_j\}$ form orthonormal bases in $\mathcal{H}_s$ and
$\mathcal{H}_t$, respectively, and $\sigma_j \ge 0$ are the non-negative real coefficients.

In order to quantify non-separability of each of these spatiotemporal modes, we define an analog of the Schmidt number denoted as $\mathcal{C}$ given by
\begin{equation}
\mathcal{C}
=
\frac{\left(\sum_j \sigma_j\right)^2}{\sum_j \sigma_j^2}.
\end{equation}

For a perfectly separable mode function, $\mathcal{C}=1$, whereas $\mathcal{C}>1$ therefore indicates non-separability or non-zero coupling between the two degrees of freedom.

\begin{figure}[H]
    \centering
    \includegraphics[width=0.6\textwidth]{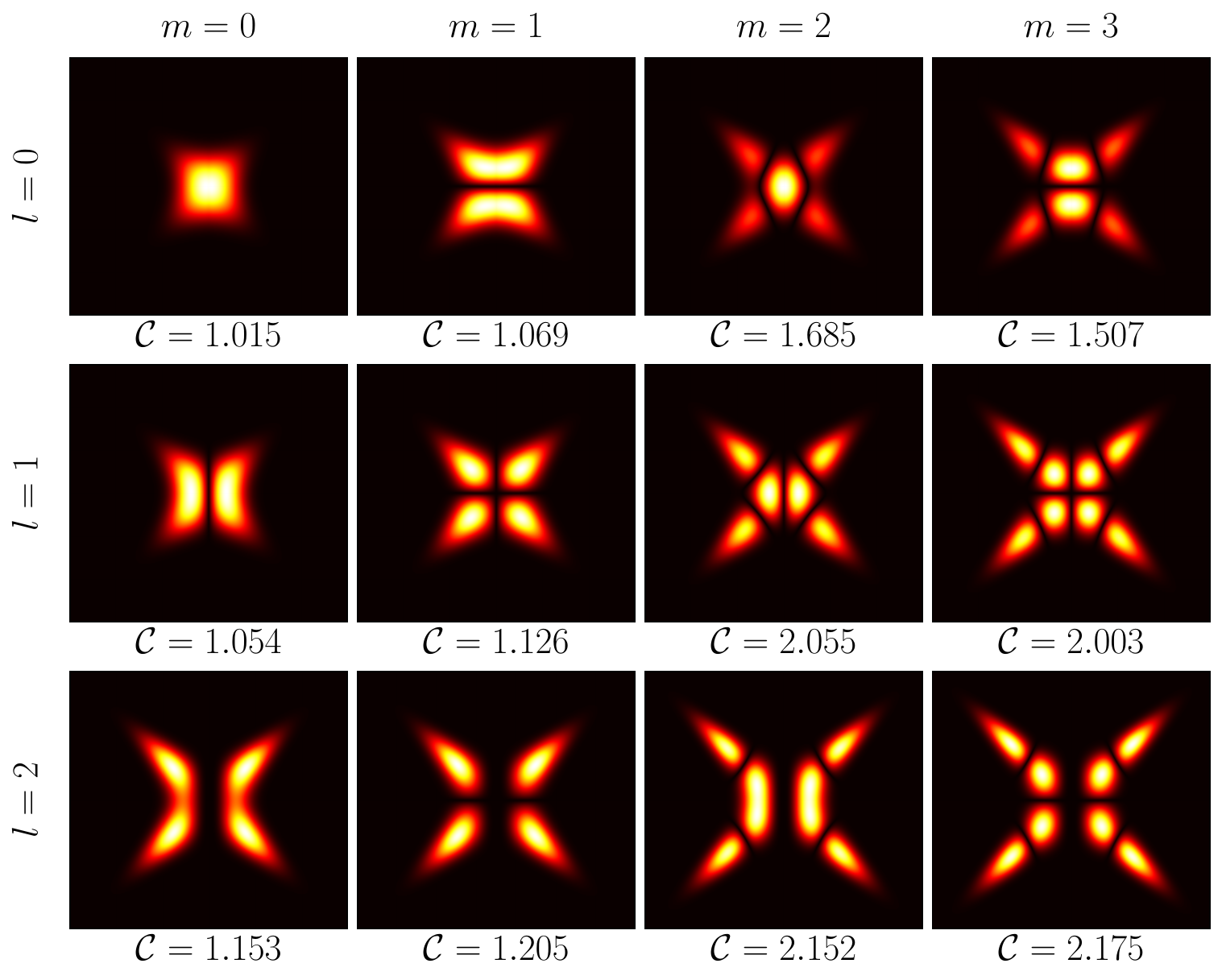}
   \caption{Spatiotemporal Schmidt modes and their degree of non-separability $\mathcal{C}$.}
    \label{smfig2}
\end{figure}

Figure~\ref{smfig2} shows the first few spatiotemporal modes and their corresponding degrees of non-separability. The fundamental mode is observed to be approximately separable, with $\mathcal{C} \approx 1$. As the mode order increases, the strength of the coupling between the spatial and temporal degrees of freedom increases, thereby causing $\mathcal{C}$ to increase.

\section{Effect of Frequency Bandwidth and Crystal Angle on the Schmidt Number}

\begin{figure}[H]
    \centering
    \vspace{-10pt} 
    \includegraphics[width=0.99\textwidth]{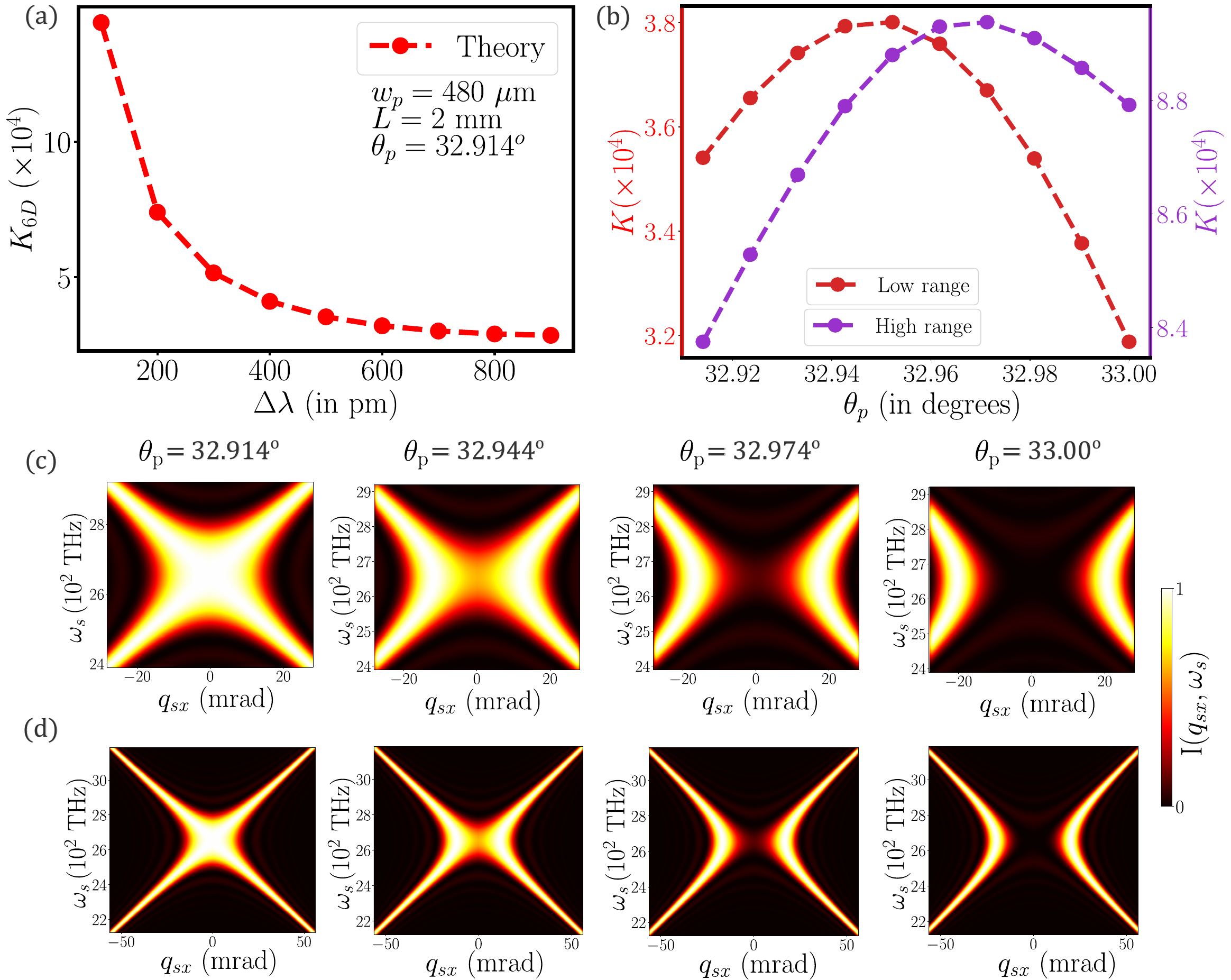}
    \vspace{-5pt}  
    \caption{Variation of the Schmidt number with (a) frequency bandwidth and (b) crystal angle. In (b), the red curve shows the Schmidt number obtained using a smaller numerical integration range, while the violet curve corresponds to a calculation performed with twice that range. Figure (c) and (d) show the corresponding near-field intensity profiles for the smaller and larger numerical ranges, respectively, for increasing $\theta_{p}$ from $\theta_{p}=32.914^\circ$ corresponding to collinear emission to $\theta_{p}=33.00^\circ$ corresponding to non-collinear emission of the degenerate signal and idler photons.
}
    \label{fig:smfig3}
    \vspace{-10pt} 
\end{figure}

Figure~\ref{fig:smfig3}(a) shows that the degree of spatiotemporal entanglement as quantified by the Schmidt number $K$ decreases with increasing wavelength bandwidth. We also observe that the Schmidt number $k$ increases with the beam waist as shown in Fig~(2)(b) of the main paper. These observations indicate that the degree of entanglement is approximately inversely proportional to the pump spectral volume, defined as $\Delta \omega_{p} / w_p^2$, as discussed in Ref.~[25] of the main paper.

The Schmidt number can be understood as the ratio of the effective volume of the spatiotemporal phase-matching function and the pump spectral volume (see Eq.~(34) of Ref.~[25] of the main paper). This effective volume increases with the crystal angle while the spectral volume of the pump remains constant, resulting in an initial enhancement of the degree of entanglement. However, beyond a certain crystal angle, a portion of this effective volume extends beyond the finite numerical window used in the calculations. Consequently, the calculated Schmidt number begins to decrease with further increases in the crystal angle. This reduction arises from a geometrical limitation imposed by the finite integration range of the numerical calculation.

This effect is illustrated in Fig.~\ref{fig:smfig3}(b), where Schmidt numbers computed using two different numerical ranges are compared. The red curve corresponds to a smaller calculation window, while the violet curve corresponds to a window with twice the range of the former. The larger window captures a greater fraction of the phase-matching volume, thereby reducing the artificial decrease in the Schmidt number observed at larger crystal angles. Figures~\ref{fig:smfig3}(c) and \ref{fig:smfig3}(d) show the corresponding intensity profiles at different crystal angles $(\theta_p)$, obtained using smaller and larger numerical ranges, respectively. These intensity distributions clearly demonstrate that extending the calculation window allows more spatial and temporal extent of the phase-matching function to be captured, leading to a more accurate characterization of the degree of entanglement.